\documentclass{iau}
\usepackage{graphicx}
\usepackage{natbib}
\usepackage{enumerate}
\usepackage{amsmath}   

\usepackage[pdftitle={Constraints on the habitability of extrasolar moons}, colorlinks = true, breaklinks = true, citecolor = blue, linkcolor = blue, urlcolor = blue, pdfauthor = {Ren\'{e} Heller}]{hyperref}

\title[Constraints on the habitability of extrasolar moons] 
{Constraints on the habitability of \\ extrasolar moons}

\author[Ren\'e Heller \& Rory Barnes]   
{Ren\'e Heller$^1$
 \and Rory Barnes$^{2,3}$}

\affiliation{$^1$Leibniz Institute for Astrophysics Potsdam (AIP), An der Sternwarte 16, 14482 Potsdam \\
email: {\tt rheller@aip.de} \\[\affilskip]
$^2$University of Washington, Dept. of Astronomy, Seattle, WA 98195, USA \\[\affilskip]
$^3$Virtual Planetary Laboratory, NASA, USA \\
email: {\tt rory@astro.washington.edu}}

\pubyear{2012}
\volume{293}  
\pagerange{xxx-xxx}
\jname{Formation, detection, and characterization of extrasolar habitable planets}
\editors{Nader Haghighipour}
\begin{document}

\maketitle

\begin{abstract}
Detections of massive extrasolar moons are shown feasible with the \textit{Kepler} space telescope. \textit{Kepler}'s findings of about $50$ exoplanets in the stellar habitable zone naturally make us wonder about the habitability of their hypothetical moons. Illumination from the planet, eclipses, tidal heating, and tidal locking distinguish remote characterization of exomoons from that of exoplanets. We show how evaluation of an exomoon's habitability is possible based on the parameters accessible by current and near-future technology.
\keywords{celestial mechanics -- planets and satellites: general -- astrobiology -- eclipses}
\end{abstract}

\firstsection 
\section{Introduction}

\noindent
The possible discovery of inhabited exoplanets has motivated considerable efforts towards estimating planetary habitability. Effects of stellar radiation \citep{1993Icar..101..108K,2007A&A...476.1373S}, planetary spin \citep{1997Icar..129..254W,2009ApJ...691..596S}, tidal evolution \citep{2008MNRAS.391..237J,2009ApJ...700L..30B,2011A&A...528A..27H}, and composition \citep{2006Sci...313.1413R,2010ApJ...715.1050B} have been studied.

Meanwhile, \textit{Kepler}'s high precision has opened the possibility of detecting extrasolar moons \citep{2009MNRAS.400..398K,2011ApJ...743...97T} and the first dedicated searches for moons in the \textit{Kepler} data are underway \citep{2012ApJ...750..115K}. With the detection of an exomoon in the stellar irradiation habitable zone (IHZ) at the horizon, exomoon habitability is now drawing scientific and public attention. Yet, investigations on exomoon habitability are rare \citep{1987AdSpR...7..125R,1997Natur.385..234W,2006ApJ...648.1196S,2012arXiv1209.5323H,2012A&A...545L...8H}. These studies have shown that illumination from the planet, satellite eclipses, tidal heating, and constraints from orbital stability have fundamental effects on the habitability of moons -- at least as important as irradiation from the star. In this communication we review our recent findings of constraints on exomoon habitability.

\section{Why bother about exomoon habitability?}

\noindent
The number of confirmed exoplanets will soon run into the thousands with only a handful being located in the IHZ. Why should we bother about the habitability of their moons when it is yet so hard to characterize even the planets? We adduce four reasons:

\begin{figure}[ht]
\begin{center}
 \includegraphics[width=5.37in]{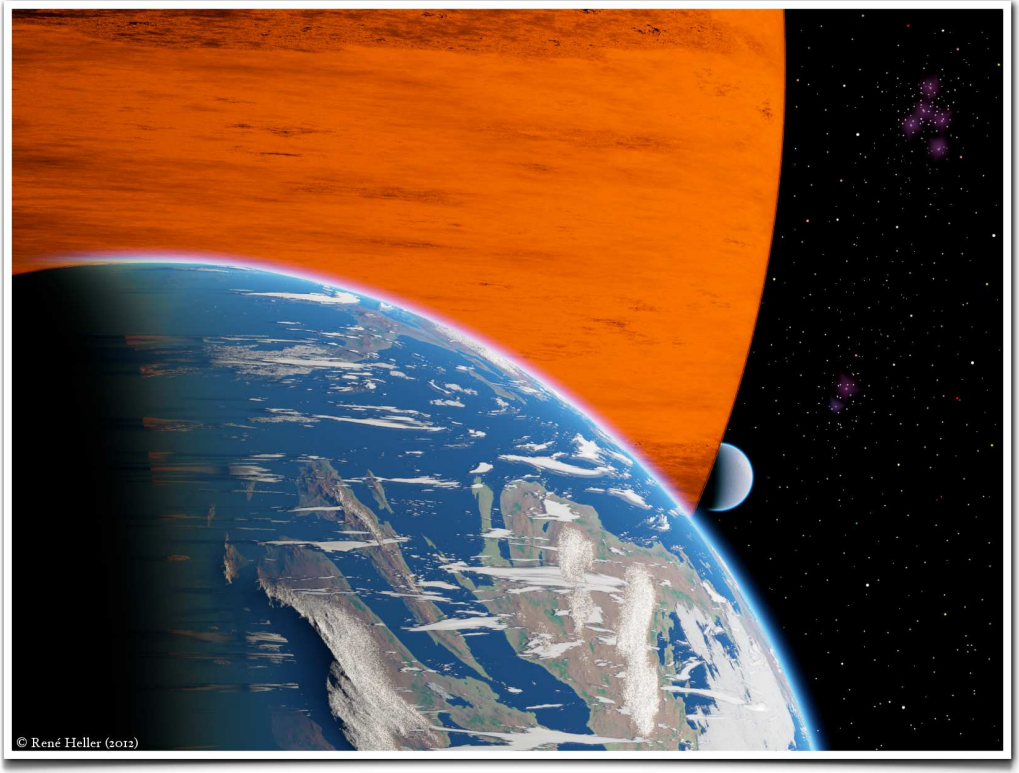} 
 \caption{A hypothetical Earth-sized moon orbiting the recently discovered Neptune-sized planet Kepler-22\,b in the irradiation habitable zone of a Sun-like host star. The satellite's orbit is equal to Europa's distance from Jupiter. A second moon with the size of Europa is in the background.}
   \label{fig:Kepler}
\end{center}
\end{figure}

\begin{enumerate}
\item[\textbf{(\textit{i.})}] \ If they exist, then the first detected exomoons will be roughly Earth-sized, i.e. have masses $\gtrsim0.25\,M_\oplus$ \citep{2009MNRAS.400..398K}.
\item[\textbf{(\textit{ii.})}] \ Expected to be tidally locked to their planets, exomoons in the IHZ have days much shorter than their stellar year. This is an advantage for their habitability compared to terrestrial planets in the IHZ of M dwarfs, which become tidally locked to the star.
\item[\textbf{(\textit{iii.})}] \ Massive host planets of satellites are more likely to maintain their primordial spin-orbit misalignment than small planets \citep{2011A&A...528A..27H}. Thus, an extrasolar moon in the stellar IHZ which will likely orbit any massive planet in its equatorial plane \citep{2011ApJ...736L..14P} is much more likely to experience seasons than a single terrestrial planet at the same distance from the star.
\item[\textbf{(\textit{iv.})}] \ Extrasolar habitable moons could be much more numerous than planets. In her IAU talk on Aug. 28, Natalie Batalha has shown the ``Periodic Table of Exoplanets'', indicating many more Warm Neptunes and Warm Jovians (i.e. potential hosts to habitable moons) than Warm Earths in the \textit{Kepler} data (\url{http://phl.upr.edu}).
\end{enumerate}

The confirmation of the Neptune-sized planet Kepler-22\,b in the IHZ of a Sun-like star \citep{2012ApJ...745..120B} and the detection of Kepler-47\,c in the IHZ of a stellar binary system \citep{2012Sci...337.1511O} have shown that, firstly, adequate host planets exist and, secondly, their characterization is possible. Figure~\ref{fig:Kepler} displays a hypothetical, inhabited Earth-sized moon about Kepler-22\,b in an orbit as wide as Europa's semi-major axis about Jupiter.

\section{Constraints on exomoon habitability}

\begin{figure}[b]
 \includegraphics[width=5.31in]{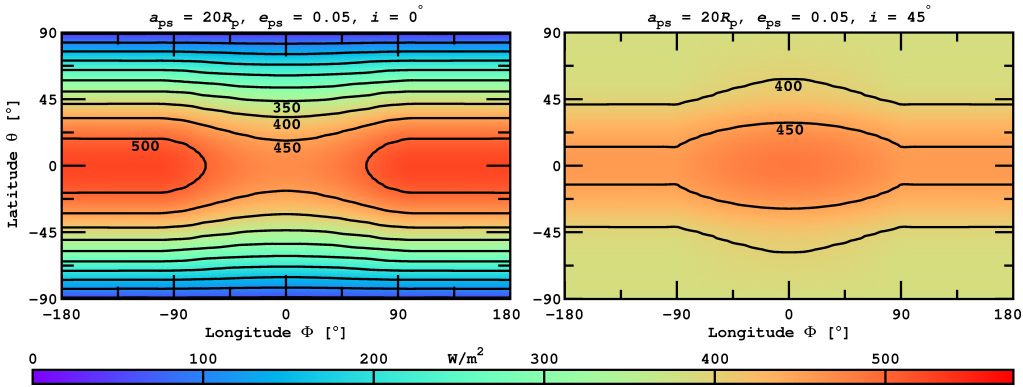} 
 \caption{Orbit-averaged surface illumination of a hypothetical exomoon orbiting Kepler-22\,b. Stellar reflection and thermal emission from the planet as well as tidal heating are included. \textit{Left}: Orbital inclination is $0^\circ$, i.e. the moon is subject to periodic eclipses. The subplanetary point at ($\phi=0^\circ=\theta$) is the coldest location. \textit{Right}: Same configuration as in the left panel, except for an inclination of $45^\circ$. Eclipses are rare and the planet's illumination makes the subplanetary point the warmest location on the moon.}
   \label{fig:illumination}
\end{figure}

\subsection{Illumination}

\noindent
Similar to the case of planets, where the possibility of liquid surface water defines habitability \citep{1993Icar..101..108K}, we can approach a satellite's habitability by estimating its orbit-averaged global energy flux $\bar{F}_\mathrm{s}^\mathrm{glob}$. If this top-of-the-atmosphere quantity is less than the critical flux to induce a runaway greenhouse process $F_\mathrm{RG}$ and if the planet-moon duet is in the IHZ, then the moon can be considered habitable. Of course, a planet-moon system can also orbit a star outside the IHZ and tidal heating could prevent the moon from becoming a snowball \citep{2006ApJ...648.1196S}. However, the geophysical and atmospheric properties of extremely tidally-heated bodies are unknown, making habitability assessments challenging. With Io's surface tidal heating of about $2\,\mathrm{W/m}^2$ \citep{2000Sci...288.1198S} in mind, which leads to rapid reshaping of the moon's surface and global volcanism, we thus focus on moons in the IHZ for the time being.

Computation of $\bar{F}_\mathrm{s}^\mathrm{glob}$ includes phenomena that are mostly irrelevant for planets. We must consider the planet's stellar reflection and thermal emission as well as tidal heating in the moon. Only in wide circular orbits these effects will be negligible. Let us assume a hypothetical moon about Kepler-22\,b, which is tidally locked to the planet. In Fig.~\ref{fig:illumination} we show surface maps of its flux averaged over one stellar orbit and for two different orbital configurations. In both scenarios the satellite's semi-major axis is $20$ planetary radii and eccentricity is $0.05$. Tidal surface heating, assumed to be distributed uniformly over the surface, is $0.017\,\mathrm{W/m}^2$ in both panels. For reference, the Earth's outward heat flow is $0.065\,\mathrm{W/m}^2$ through the continents and $0.1\,\mathrm{W/m}^2$ through the ocean crust \citep{2007SSRv..129...35Z}. Parametrization of the star-planet system follows \citet{2012ApJ...745..120B}. In the left panel, the moon's orbit about the planet is assumed to be coplanar with the circumstellar orbit, i.e. inclination $i=0^\circ$. The satellite is subject to eclipses almost once per orbit about the planet. An observer on the moon could only watch eclipses from the hemisphere which is permanently faced towards the planet, i.e. the moon's pro-planetary hemisphere. Eclipses are most prominent on the sub-planetary point and make it the coldest point on the moon in terms of average illumination \citep{2012arXiv1209.5323H}. In the right panel, the moon's orbit is tilted by $45^\circ$ against the circumstellar orbit and eclipses occur rarely \citep[for satellite eclipses see Fig.~1 in][]{2012A&A...545L...8H}. Illumination from the planet overcompensates for the small reduction of stellar illumination and makes the sub-planetary point the warmest spot on the moon.

To quantify a moon's habitability we need to know its average energy flux. In \citet{2012arXiv1209.5323H}Ê\ and \citet{2012A&A...545L...8H} \ we show that 

\begin{align} \label{eq:F_glob} \nonumber
\bar{F}_\mathrm{s}^\mathrm{glob} &= \frac{L_* \ (1-\alpha_\mathrm{s})}{16{\pi}a_\mathrm{*p}^2\sqrt{1-e_\mathrm{*p}^2}}
                    {\Bigg (} x_\mathrm{s} + \frac{{\pi}R_\mathrm{p}^2\alpha_\mathrm{p}}{2a_\mathrm{ps}^2} {\Bigg )} \\
             & \ \ \ \ + \frac{R_\mathrm{p}^2 \sigma_\mathrm{SB} (T_\mathrm{p}^\mathrm{eq})^4}{a_\mathrm{ps}^2} \frac{(1-\alpha_\mathrm{s})}{4} + h_\mathrm{s}(e_\mathrm{ps},a_\mathrm{ps},R_\mathrm{s}) \ \ \ , \tag{1}
\end{align}

\noindent
where $L_*$ is stellar luminosity, $a_\mathrm{*p}$ the semi-major axis of the planet's orbit about the star, $a_\mathrm{ps}$ the semi-major axis of the satellite's orbit about the planet, $e_\mathrm{*p}$ the circumstellar orbital eccentricity, $R_\mathrm{p}$ the planetary radius, $\alpha_\mathrm{p}$ and $\alpha_\mathrm{s}$ are the albedos of the planet and the satellite, respectively, $T_\mathrm{p}^\mathrm{eq}$ is the planet's thermal equilibrium temperature, $h_\mathrm{s}$ the satellite's surface-averaged tidal heating flux, $\sigma_\mathrm{SB}$ the Stefan-Boltzmann constant, and $x_\mathrm{s}$ is the fraction of the satellite's orbit that is \textit{not} spent in the shadow of the planet. Note that tidal heating $h_\mathrm{s}$ depends on the satellite's orbital eccentricity $e_\mathrm{ps}$, its semi-major axis $a_\mathrm{ps}$, and on its radius $R_\mathrm{s}$.

This formula is valid for any planetary eccentricity; it includes decrease of average stellar illumination due to eclipses; it considers stellar reflection from the planet; it accounts for the planet's thermal radiation; and it adds tidal heating. Analyses of a planet's transit timing \citep{1999A&AS..134..553S,2006A&A...450..395S} \& transit duration \citep{2009MNRAS.392..181K,2009MNRAS.396.1797K} variations in combination with direct observations of the satellite transit \citep{2006A&A...450..395S,2007A&A...470..727S,2011ApJ...743...97T} can give reasonable constraints on Eq.~\eqref{eq:F_glob} and thus on a moon's habitability. In principle, these data could be obtained with \textit{Kepler} observations alone but $N$-body simulations including tidal dissipation will give stronger constraints on the satellite's eccentricity than observations.

\subsection{The habitable edge and Hill stability}

The range of habitable orbits about a planet in the IHZ is limited by an outer and an inner orbit. The widest possible orbit is given by the planet's sphere of gravitational dominance, i.e. Hill stability, the innermost orbit is defined by the runaway greenhouse limit $\bar{F}_\mathrm{s}^\mathrm{glob}=F_\mathrm{RG}$ and is called the ``habitable edge'' \citep{2012arXiv1209.5323H}.

\begin{figure}[t]
 \hspace{.55in}
 \includegraphics[width=3.3in]{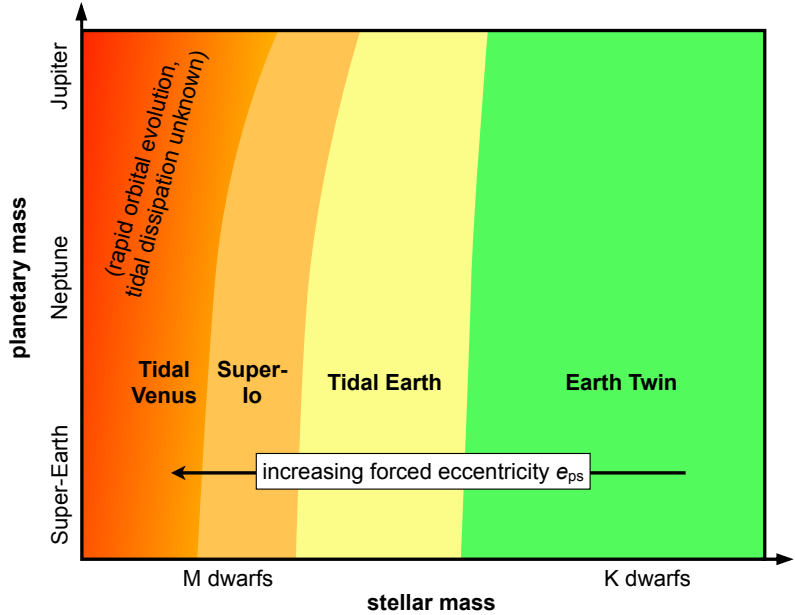} 
 \caption{Schematic classification of hypothetical $10\,M_\mathrm{Gan}$-mass ($\approx0.25\,M_\oplus$) moons in the widest Hill stable orbits about planets in the stellar IHZ. In the IHZ about M dwarfs a satellite's eccentricity $e_\mathrm{ps}$ is strongly forced by the close star, which induces strong tidal heating in the moon. A Tidal Venus moon is uninhabitable.}
   \label{fig:exomoon_flavors}
\end{figure}

Space for habitable orbits about a planet decreases when the planet's Hill radius moves inward and when the habitable edge moves outward. This is what happens when we virtually move a given planet-moon binary from the IHZ of G a star to that of a K star and finally into the IHZ of an M star \citep{2012A&A...545L...8H}. Shrinking the planet's Hill sphere means that any moon must orbit the planet ever closer to remain gravitationally bound. Additionally, perturbations from the star on the moon's orbit become substantial and due to the enhanced eccentricity and the accompanying tidal heating the habitable edge moves outward. Hence, the range of habitable orbits vanishes. As the planetary Hill sphere in the IHZ about an M dwarf is small and moons follow eccentric orbits, satellites in M dwarf systems become subject to catastrophic tidal heating, and this energy dissipation induces rapid evolution of their orbits.

Let us take an example: Imagine a planet-moon binary composed of a Neptune-sized host and a satellite $10$ times the mass of Ganymede ($10\,M_\mathrm{Gan}\approx0.25\,M_\oplus$, $M_\oplus$ being the mass of the Earth). This duet shall orbit in the center of the IHZ of a $0.5\,M_\odot$ star ($M_\odot$ being the solar mass), i.e. at a stellar distance of roughly $0.3$\,AU \citep{2007A&A...476.1373S}. The outermost stable satellite orbit, which numerical simulations have shown to be generally about $1/3$ the planet's Hill radius \citep{2002ApJ...575.1087B}, then turns out at $486,000$\,km. This means that the moon's circum-planetary orbit must be at least as tight as Io's orbit about Jupiter! Recall that Io is subject to enormous tidal heating. Yet, the true tidal heating of this hypothetical moon will depend on its eccentricity (potentially forced by the close star and/or by further satellites), and with masses and radii different from those of Jupiter and Io, tidal dissipation in that system will be different.

Moving on to a $0.25\,M_\odot$ star, the IHZ is now at $\approx0.125$\,AU and the satellite's orbit must be within $255,000$\,km about the planet. For a $0.1\,M_\odot$ star with its IHZ at about $0.05$\,AU the moon's orbital semi-major axis must be $<139,000$\,km, i.e. almost as close as Miranda's orbit about Uranus. As we virtually decrease the stellar mass and as we move our planet-moon binary towards the star to remain the IHZ, the star also forces the satellite's orbit to become more and more eccentric. We expect that for stellar masses below about $0.2\,M_\odot$ no habitable Super-Ganymede exomoon can exist in the stellar IHZ due strong tidal dissipation \citep{2012A&A...545L...8H}.

Figure~\ref{fig:exomoon_flavors} shall illustrate our gedankenexperiment. The abscissae indicates stellar mass, the ordinate denotes planetary mass. For each star-planet system the planet-moon binary is assumed to orbit in the middle of the IHZ and the moon shall orbit at the widest possible orbit from the planet. With this conservative assumption, tidal heating is minimized. Colored areas indicate the type of planet according to our classification scheme proposed in \citet{BarnesHeller12}. Exomoons in K dwarf systems will hardly be subject to the dynamical constraints illustrated above (green area), thus Earth twin satellites could exist. However, if roughly Earth-mass exomoons exist in lower-mass stellar systems, then they can only occur as Tidal Earths with small but significant tidal heating (yellow area); as Super-Ios with heating $>~2\,\mathrm{W/m}^2$ but not enough to induce a runaway greenhouse process (orange area); or Tidal Venuses, i.e. with powerful tides and $\bar{F}_\mathrm{s}^\mathrm{glob}>F_\mathrm{RG}$ (red areas). A Tidal Venus is uninhabitable by definition. Tidal dissipation in the upper-left corner of Fig.~\ref{fig:exomoon_flavors} will be enormous and will lead to so far unexplored geological and orbital evolution on short timescales.

\section{Prospects for habitable extrasolar moons}

\noindent
The quest of habitable moons seeks objects unknown from the solar system. Even the most massive moon, Ganymede, has a mass of only $\approx0.025\,M_\oplus$. It is not clear whether moons as massive as Mars ($\approx0.1\,M_\oplus$) or $10$ times as massive as Ganymede ($\approx0.25\,M_\oplus$) exist \citep{2010ApJ...714.1052S,2012ApJ...753...60O}. But given the unexpected presence of giant planets orbiting their stars in only a few days and given transiting planetary systems about binary stellar systems, clearly a Mars-sized moon about a Neptune-mass planet does not sound absurd.

Although Fig.~\ref{fig:exomoon_flavors} is schematic and urgently requires deeper investigations, it indicates that habitable exomoons cannot exist in the IHZ of stars with masses $\lesssim0.2\,M_\odot$ \citep{2012A&A...545L...8H}. Orbital simulations, eventually coupled with atmosphere or climate models, have yet to be done to quantify these constraints. With NASA's \textit{James Webb Space Telescope} and ESO's \textit{European Extremely Large Telescope} facilities capable of tracking spectral signatures from inhabited exomoon are being built \citep{2010ApJ...712L.125K} and future observers will need a priority list of the most promising targets to host extraterrestrial life. Exomoons have the potential to score high if their habitability can be constrained from both high-quality observations and orbital simulations.

\bibliographystyle{aa} 
\bibliography{2012-5_HellerBarnes--IAU_Proceedings}

\begin{thebibliography}{32}
\expandafter\ifx\csname natexlab\endcsname\relax\def\natexlab#1{#1}\fi

\bibitem[{{Barnes} \& {O'Brien}(2002)}]{2002ApJ...575.1087B}
{Barnes}, J.~W. \& {O'Brien}, D.~P. 2002, \textit{ApJ}, 575, 1087

\bibitem[{{Barnes} \& {Heller}(2012)}]{BarnesHeller12}
{Barnes}, R. \& {Heller}, R. 2012, \textit{Astrobiology} (submitted)

\bibitem[{{Barnes} {et~al.}(2009){Barnes}, {Jackson}, {Greenberg}, \&
  {Raymond}}]{2009ApJ...700L..30B}
{Barnes}, R., {Jackson}, B., {Greenberg}, R., \& {Raymond}, S.~N. 2009,
  \textit{ApJL}, 700, L30

\bibitem[{{Bond} {et~al.}(2010){Bond}, {O'Brien}, \&
  {Lauretta}}]{2010ApJ...715.1050B}
{Bond}, J.~C., {O'Brien}, D.~P., \& {Lauretta}, D.~S. 2010, \textit{ApJ}, 715,
  1050

\bibitem[{{Borucki} {et~al.}(2012){Borucki}, {Koch}, {Batalha}, {Bryson},
  {Rowe}, {Fressin}, {Torres}, {Caldwell}, {Christensen-Dalsgaard}, {Cochran},
  {DeVore}, {Gautier}, {Geary}, {Gilliland}, {Gould}, {Howell}, {Jenkins},
  {Latham}, {Lissauer}, {Marcy}, {Sasselov}, {Boss}, {Charbonneau}, {Ciardi},
  {Kaltenegger}, {Doyle}, {Dupree}, {Ford}, {Fortney}, {Holman}, {Steffen},
  {Mullally}, {Still}, {Tarter}, {Ballard}, {Buchhave}, {Carter},
  {Christiansen}, {Demory}, {D{\'e}sert}, {Dressing}, {Endl}, {Fabrycky},
  {Fischer}, {Haas}, {Henze}, {Horch}, {Howard}, {Isaacson}, {Kjeldsen},
  {Johnson}, {Klaus}, {Kolodziejczak}, {Barclay}, {Li}, {Meibom}, {Prsa},
  {Quinn}, {Quintana}, {Robertson}, {Sherry}, {Shporer}, {Tenenbaum},
  {Thompson}, {Twicken}, {Van Cleve}, {Welsh}, {Basu}, {Chaplin}, {Miglio},
  {Kawaler}, {Arentoft}, {Stello}, {Metcalfe}, {Verner}, {Karoff}, {Lundkvist},
  {Lund}, {Handberg}, {Elsworth}, {Hekker}, {Huber}, {Bedding}, \&
  {Rapin}}]{2012ApJ...745..120B}
{Borucki}, W.~J., {Koch}, D.~G., {Batalha}, N., {et~al.} 2012, \textit{ApJ},
  745, 120

\bibitem[{{Heller}(2012)}]{2012A&A...545L...8H}
{Heller}, R. 2012, \textit{A\&A}, 545, L8

\bibitem[{{Heller} \& {Barnes}(2012)}]{2012arXiv1209.5323H}
{Heller}, R. \& {Barnes}, R. 2012, \textit{Astrobiology} (in press), ArXiv
  e-prints

\bibitem[{{Heller} {et~al.}(2011){Heller}, {Leconte}, \&
  {Barnes}}]{2011A&A...528A..27H}
{Heller}, R., {Leconte}, J., \& {Barnes}, R. 2011, \textit{A\&A}, 528, A27

\bibitem[{{Jackson} {et~al.}(2008){Jackson}, {Barnes}, \&
  {Greenberg}}]{2008MNRAS.391..237J}
{Jackson}, B., {Barnes}, R., \& {Greenberg}, R. 2008, \textit{MNRAS}, 391, 237

\bibitem[{{Kaltenegger}(2010)}]{2010ApJ...712L.125K}
{Kaltenegger}, L. 2010, \textit{ApJL}, 712, L125

\bibitem[{{Kasting} {et~al.}(1993){Kasting}, {Whitmire}, \&
  {Reynolds}}]{1993Icar..101..108K}
{Kasting}, J.~F., {Whitmire}, D.~P., \& {Reynolds}, R.~T. 1993,
  \textit{Icarus}, 101, 108

\bibitem[{{Kipping}(2009{\natexlab{a}})}]{2009MNRAS.392..181K}
{Kipping}, D.~M. 2009{\natexlab{a}}, \textit{MNRAS}, 392, 181

\bibitem[{{Kipping}(2009{\natexlab{b}})}]{2009MNRAS.396.1797K}
{Kipping}, D.~M. 2009{\natexlab{b}}, \textit{MNRAS}, 396, 1797

\bibitem[{{Kipping} {et~al.}(2012){Kipping}, {Bakos}, {Buchhave},
  {Nesvorn{\'y}}, \& {Schmitt}}]{2012ApJ...750..115K}
{Kipping}, D.~M., {Bakos}, G.~{\'A}., {Buchhave}, L., {Nesvorn{\'y}}, D., \&
  {Schmitt}, A. 2012, \textit{ApJ}, 750, 115

\bibitem[{{Kipping} {et~al.}(2009){Kipping}, {Fossey}, \&
  {Campanella}}]{2009MNRAS.400..398K}
{Kipping}, D.~M., {Fossey}, S.~J., \& {Campanella}, G. 2009, \textit{MNRAS},
  400, 398

\bibitem[{{Ogihara} \& {Ida}(2012)}]{2012ApJ...753...60O}
{Ogihara}, M. \& {Ida}, S. 2012, \textit{ApJ}, 753, 60

\bibitem[{{Orosz} {et~al.}(2012){Orosz}, {Welsh}, {Carter}, {Fabrycky},
  {Cochran}, {Endl}, {Ford}, {Haghighipour}, {MacQueen}, {Mazeh},
  {Sanchis-Ojeda}, {Short}, {Torres}, {Agol}, {Buchhave}, {Doyle}, {Isaacson},
  {Lissauer}, {Marcy}, {Shporer}, {Windmiller}, {Barclay}, {Boss}, {Clarke},
  {Fortney}, {Geary}, {Holman}, {Huber}, {Jenkins}, {Kinemuchi}, {Kruse},
  {Ragozzine}, {Sasselov}, {Still}, {Tenenbaum}, {Uddin}, {Winn}, {Koch}, \&
  {Borucki}}]{2012Sci...337.1511O}
{Orosz}, J.~A., {Welsh}, W.~F., {Carter}, J.~A., {et~al.} 2012,
  \textit{Science}, 337, 1511

\bibitem[{{Porter} \& {Grundy}(2011)}]{2011ApJ...736L..14P}
{Porter}, S.~B. \& {Grundy}, W.~M. 2011, \textit{ApJL}, 736, L14

\bibitem[{{Raymond} {et~al.}(2006){Raymond}, {Mandell}, \&
  {Sigurdsson}}]{2006Sci...313.1413R}
{Raymond}, S.~N., {Mandell}, A.~M., \& {Sigurdsson}, S. 2006, \textit{Science},
  313, 1413

\bibitem[{{Reynolds} {et~al.}(1987){Reynolds}, {McKay}, \&
  {Kasting}}]{1987AdSpR...7..125R}
{Reynolds}, R.~T., {McKay}, C.~P., \& {Kasting}, J.~F. 1987, \textit{Advances
  in Space Research}, 7, 125

\bibitem[{{Sartoretti} \& {Schneider}(1999)}]{1999A&AS..134..553S}
{Sartoretti}, P. \& {Schneider}, J. 1999, \textit{A\&AS}, 134, 553

\bibitem[{{Sasaki} {et~al.}(2010){Sasaki}, {Stewart}, \&
  {Ida}}]{2010ApJ...714.1052S}
{Sasaki}, T., {Stewart}, G.~R., \& {Ida}, S. 2010, \textit{ApJ}, 714, 1052

\bibitem[{{Scharf}(2006)}]{2006ApJ...648.1196S}
{Scharf}, C.~A. 2006, \textit{ApJ}, 648, 1196

\bibitem[{{Selsis} {et~al.}(2007){Selsis}, {Kasting}, {Levrard}, {Paillet},
  {Ribas}, \& {Delfosse}}]{2007A&A...476.1373S}
{Selsis}, F., {Kasting}, J.~F., {Levrard}, B., {et~al.} 2007, \textit{A\&A},
  476, 1373

\bibitem[{{Simon} {et~al.}(2007){Simon}, {Szatm{\'a}ry}, \&
  {Szab{\'o}}}]{2007A&A...470..727S}
{Simon}, A., {Szatm{\'a}ry}, K., \& {Szab{\'o}}, G.~M. 2007, \textit{A\&A},
  470, 727

\bibitem[{{Spencer} {et~al.}(2000){Spencer}, {Rathbun}, {Travis}, {Tamppari},
  {Barnard}, {Martin}, \& {McEwen}}]{2000Sci...288.1198S}
{Spencer}, J.~R., {Rathbun}, J.~A., {Travis}, L.~D., {et~al.} 2000, Science,
  288, 1198

\bibitem[{{Spiegel} {et~al.}(2009){Spiegel}, {Menou}, \&
  {Scharf}}]{2009ApJ...691..596S}
{Spiegel}, D.~S., {Menou}, K., \& {Scharf}, C.~A. 2009, \textit{ApJ}, 691, 596

\bibitem[{{Szab{\'o}} {et~al.}(2006){Szab{\'o}}, {Szatm{\'a}ry}, {Div{\'e}ki},
  \& {Simon}}]{2006A&A...450..395S}
{Szab{\'o}}, G.~M., {Szatm{\'a}ry}, K., {Div{\'e}ki}, Z., \& {Simon}, A. 2006,
  \textit{A\&A}, 450, 395

\bibitem[{{Tusnski} \& {Valio}(2011)}]{2011ApJ...743...97T}
{Tusnski}, L.~R.~M. \& {Valio}, A. 2011, \textit{ApJ}, 743, 97

\bibitem[{{Williams} \& {Kasting}(1997)}]{1997Icar..129..254W}
{Williams}, D.~M. \& {Kasting}, J.~F. 1997, \textit{Icarus}, 129, 254

\bibitem[{{Williams} {et~al.}(1997){Williams}, {Kasting}, \&
  {Wade}}]{1997Natur.385..234W}
{Williams}, D.~M., {Kasting}, J.~F., \& {Wade}, R.~A. 1997, \textit{Nature},
  385, 234

\bibitem[{{Zahnle} {et~al.}(2007){Zahnle}, {Arndt}, {Cockell}, {Halliday},
  {Nisbet}, {Selsis}, \& {Sleep}}]{2007SSRv..129...35Z}
{Zahnle}, K., {Arndt}, N., {Cockell}, C., {et~al.} 2007, Space Science Reviews,
  129, 35

\end{thebibliography}

\end{document}